\documentclass{article}
\usepackage{url}
\usepackage{spconf, graphicx}
\usepackage{cite}
\usepackage{subcaption}
\usepackage{amssymb}
\usepackage{academicons}

\usepackage{graphicx}

\begin{document}
	
	\title{Masked Acoustic Unit for Mispronunciation Detection and Correction}
	\name{Zhan Zhang, Yuehai Wang, and Jianyi Yang}
	\address{Department of Information and Electronic Engineering, Zhejiang University, Zhejiang
		310007, China}	

	\maketitle
	
	\begin{abstract}
		Computer-Assisted Pronunciation Training (CAPT) plays an important role in language learning. Conventional ASR-based CAPT methods require expensive annotation of the ground truth pronunciation for the supervised training. Meanwhile, certain undefined non-native phonemes cannot be correctly classified into standard phonemes, making the annotation process challenging and subjective. On the other hand, ASR-based CAPT methods only give the learner text-based feedback about the mispronunciation, but cannot teach the learner how to pronounce the sentence correctly. To solve these limitations, we propose to use the acoustic unit (AU) as the intermediary feature for both mispronunciation detection and correction. The proposed method uses the masked AU sequence and the target phonemes to detect the error AU and then corrects it.  This method can give the learner speech-based self-imitating feedback, making our CAPT powerful for education.
	\end{abstract}
	

	\section{Introduction}
	\label{sec:intro}
	Computer-Assisted Pronunciation Training (CAPT) is an important technology to offer a flexible education service for the second language (L2) learners. In the pronunciation training process, the learner is asked to read a target text, and the CAPT system should detect mispronunciations of the speech and give the proper feedback. The common approach for CAPT is to compare the pronounced speech with a certain standard pronunciation distribution. If the deviation is too large, this pronunciation is judged as an error. 
	
\begin{figure}[!t]
	\centering
	\begin{minipage}[b]{0.4\linewidth}
		\includegraphics[width=1\linewidth]{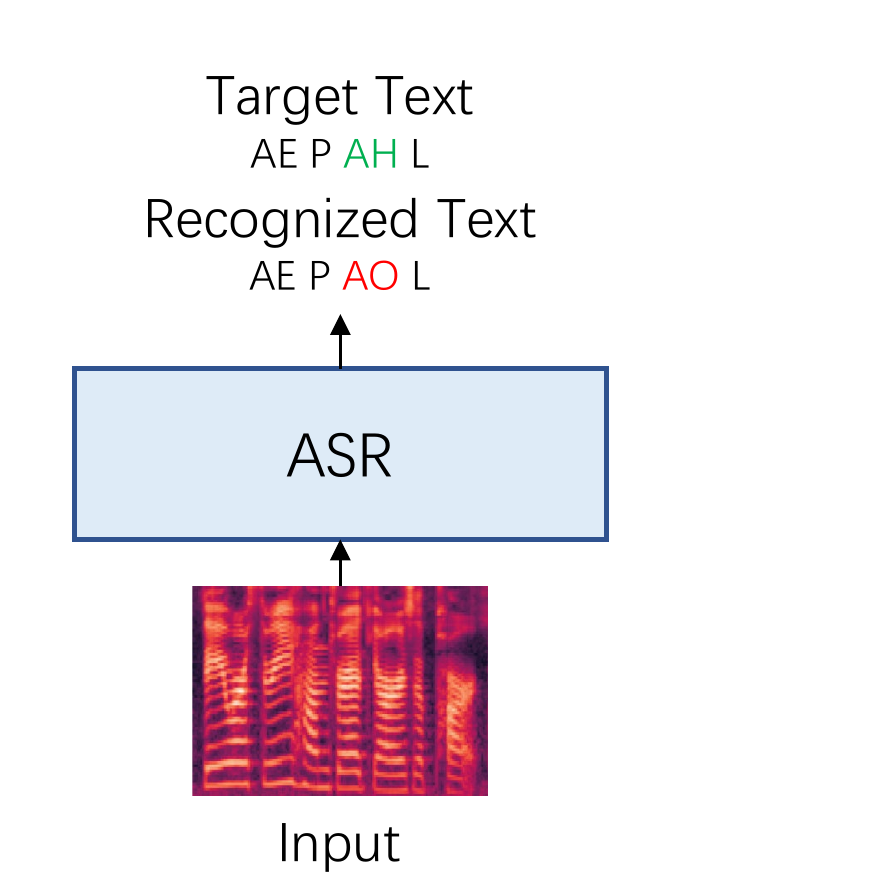}
		\subcaption{}
		\label{fig:asr_based}
	\end{minipage}
	\begin{minipage}[b]{0.4\linewidth}
		\includegraphics[width=1\linewidth]{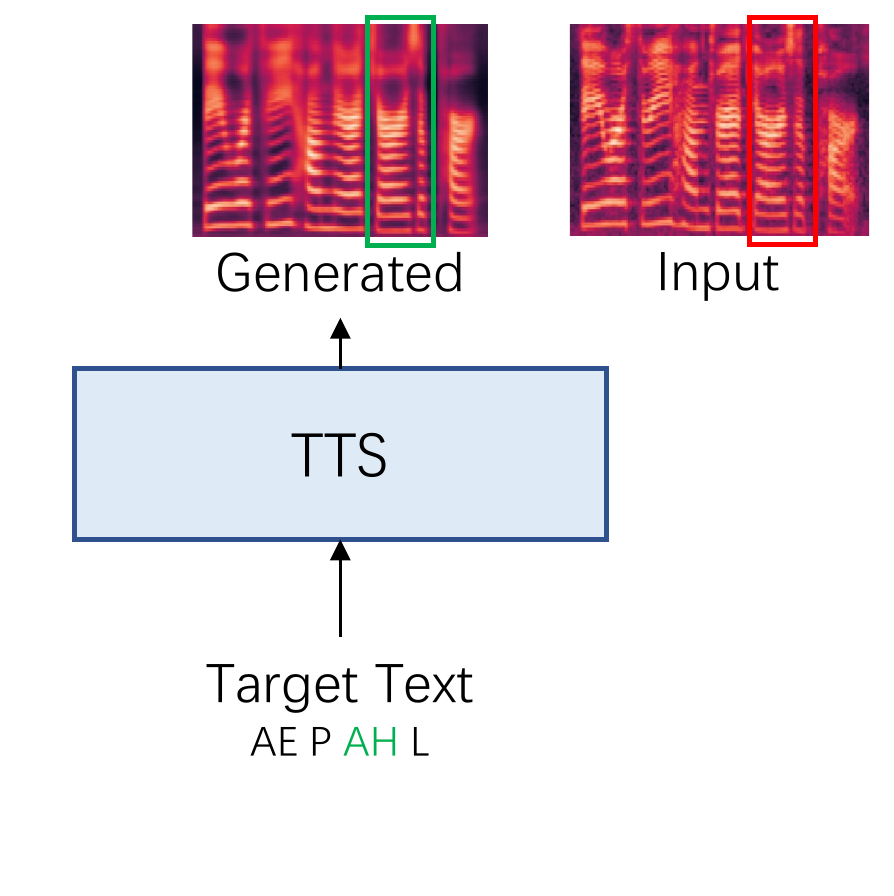}
		\subcaption{}
		\label{fig:com_based}
	\end{minipage}
	\begin{minipage}[b]{0.5\linewidth}
		\includegraphics[width=1\linewidth]{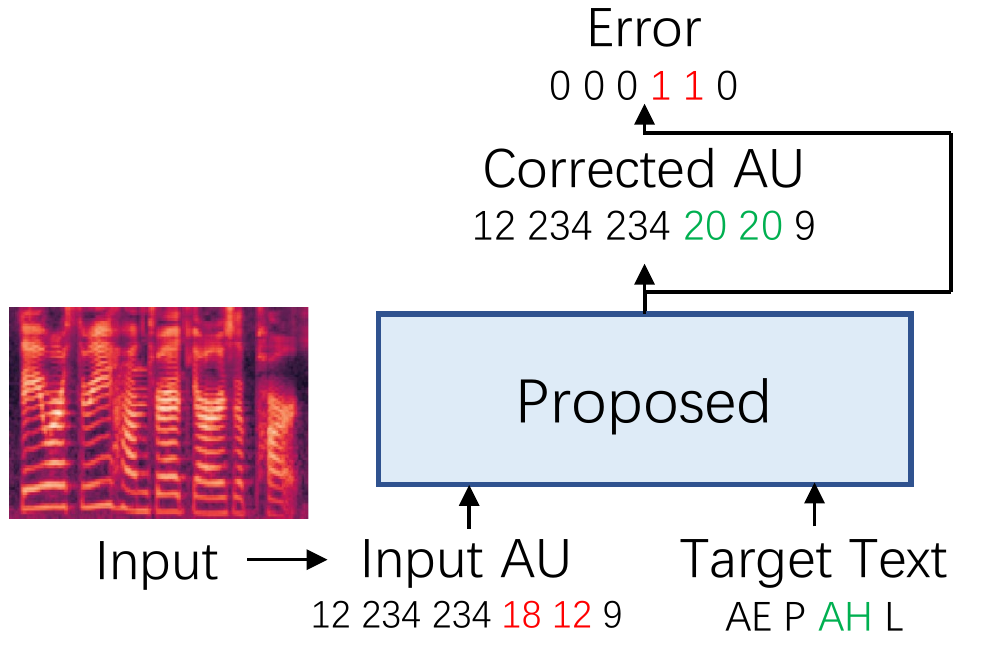}
		\subcaption{}
		\label{fig:proposed}
	\end{minipage}	
	\caption{A comparison between different CAPT methods. We mark the target feature as green and the mismatched part as red. (a) ASR-based CART systems recognize the input spectrum and align the text with the target one to find the mispronunciation. (b) Comparison-based methods can use TTS to generate the standard spectrum and find the area with high error. (c) The proposed method corrects the input AU conditioned on the target text and outputs the error prediction.}
	\label{fig:summary}
	
\end{figure}

	Currently, most CAPT systems are based upon automatic speech recognition (ASR) models. For example, the goodness-of-pronunciation (GOP)\cite{Witt2000} uses the posterior probability of the target text to judge the correctness. To simplify the workflow, \cite{Zhang2020,Yan2020,Zhang2021,Leung2019} also propose to recognize the pronounced phonemes and compare them to the target ones as shown in Fig.\ref{fig:asr_based}. 
	
	However, it is hard for these discriminative approaches to model  L2 features. First, to utilize L2 utterances, the text annotations must be the canonical phonemes pronounced by the speaker rather than the phonemes of the target text. Such annotations are expensive to obtain. Second, ASR-based methods only use the standard phonemes (for example, the 44 phonemes in English) for comparison. 
	However, as analyzed in \cite{Chang2010,Tu2018}, English learners from different language backgrounds may show acoustic characteristics similar to their mother tongues. Thus, L2 utterances may include undefined phonemes and cannot be properly classified into the restricted standard phonemes. This phenomenon also makes the phonetic annotation process challenging and subjective.
	
	To alleviate the restriction, Anti-Phone\cite{Yan2020} proposes to extend the phonemes by using augmented labels. However, it still needs the phoneme-level annotation. Recently, several unsupervised pretraining technologies\cite{Baevski2020620,Hsu9414460,vanNiekerk2020} have been applied to speech and they can discover the basic acoustic unit (AU) that is highly related to phonemes\cite{Baevski2020620}. Compared with artificially defined phonemes, AU is more fine-grained and can model the acoustic feature in an unsupervised manner. Nevertheless, how to utilize AU for mispronunciation detection still remains to be explored.
	
	On the other hand, existing ASR-based CAPT methods only give the learner text-based feedback about the mispronunciation, but cannot teach the learner how to pronounce the sentence correctly. 	In contrast, early comparison-based CAPT method \cite{Lee12220121252012} chooses to align the student utterance with a prepared teacher utterance. Misalignment between these two utterances is used to detect the mispronunciation. The teacher utterance can partly guide the learner for a correct pronunciation. With the quick development of text-to-speech (TTS) technologies, the teacher utterance can also be generated by a standard TTS model for the comparison-based methods (shown in Fig.\ref{fig:com_based}).  However, the main deficiency of such CAPT systems is that the low-level feature for comparison is not robust. The error between the student and the standard teacher feature may not indicate the mispronunciation due to speaker or style variation.
	
	In this paper, we use the vector-quantized variational autoencoder (VQ-VAE)\cite{vanNiekerk2020} to encode both L1 (native) and L2 utterances into AU as the intermediary feature. We propose a novel structure to utilize the encoded AU for both mispronunciation detection and correction as shown in Fig.\ref{fig:proposed}. The proposed method can detect the mispronunciation without expensive expert annotations. Further, by masking the error area, the correct AU is generated by the model and finally converted to the correct pronunciation. As we only modify the error area and keep the correct pronunciation, the generate speech retains the style of the speaker\footnote{Samples are provided in \url{https://zju-zhan-zhang.github.io/mispronunciation-d-c/}.}. Such a self-imitating feedback is powerful for education\cite{Yang1519September2019}. 

	\section{Proposed Method}

	\begin{figure}[tb!]
		\centering
		\includegraphics[width=1\linewidth]{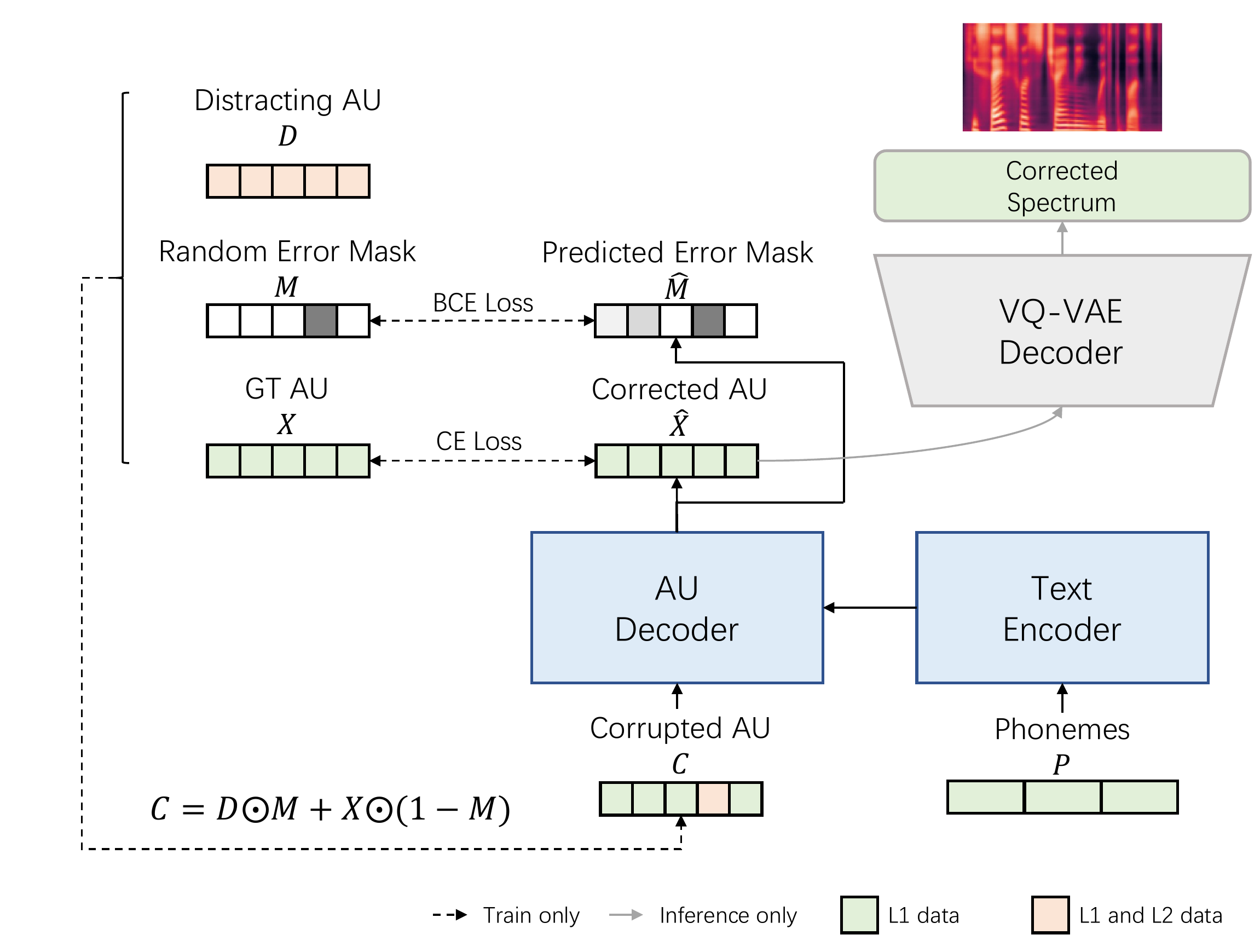}
		\caption{Workflow of the proposed method. The ground truth L1 AU sequence is corrupted by the distracting AU. The AU decoder is trained to predict the original L1 AU and the error mask based on the corrupted AU and the phonemes. For inference, the corrected AU is converted back to spectrums using the VQ-VAE decoder.}
		\label{fig:cor}
	\end{figure}
	
	\subsection{Acoustic Unit Discovery}
	We choose VQ-VAE\cite{vanNiekerk2020} as the acoustic unit discovery model, as VQ-VAE can not only encode the spectrum but also decode AU back to the spectrum compared with other speech pretraining models such as Wav2Vec\cite{Baevski2020620} or Hu-BERT\cite{Hsu9414460}. 
	
	
	Compared with the original VQ-VAE, our modified version models the spectrum instead of the raw waveform. We also replace the encoder-decoder structure to Conformer\cite{Gulati2020} as it can better capture the time-relevance.
	Moreover, to encourage the codebook usage, we append the diversity loss proposed by \cite{Baevski2020620} besides the spectrum reconstructing mean square error (MSE) loss.
	
	\subsection{Acoustic Unit Decoding}
	 Our workflow is illustrated in Fig.\ref{fig:cor}. After the training of VQ-VAE converges, we freeze the model parameters and encode both L1 and L2 utterances into AU. To simulate mispronunciations, the ground truth L1 AU sequence $X$ are mixed with the distracting AU sequence $D$ sampled from other utterances. Formally, for the L1 AU sequence $X_{0:T}$, we randomly replace $n$ segments of its original AU $X_{j_i:j_i+k_i}, 1\leq i \leq n$ 
	with the distracting AU sequence $D$, where $k$ is the replacement length and $j$ is the start position. Correspondingly, we use a mask sequence $M_{0:T}$ (its initial values are 0) and set $M_{j_i:j_i+k_i}$ to 1. The corrupted AU sequence can be denoted as
	\begin{equation}
		C=D\odot M+X\odot(1-M),
		\label{eq:mask}
	\end{equation}
	where $\odot$ is the element-wise production. 
	
	For the error detection model, as predicting the original L1 AU sequence helps the model to learn the correct distribution, we force the model to predict both the original AU sequence and the error mask based on the phonemes and the corrupted AU sequence.
	We adopt Transformer\cite{Vaswani2017} for this task. After encoding the phoneme $P$, we use two linear projection layers at the decoder for the original L1 AU and error mask prediction,
	\begin{equation}
		\hat{X},\hat{M}=\mathrm{Dec}(C,\mathrm{Enc}(P)).
	\end{equation}
	The loss function is defined as the classification loss between $\hat{X}$ and $X$, $\hat{M}$ and $M$, using the cross-entropy (CE) loss and the binary cross-entropy (BCE) loss, respectively,
	\begin{equation}
		\mathcal{L}=\mathrm{CE}(\hat{X},X)+\mathrm{BCE}(\hat{M}, M).
		\label{eg:cor}
	\end{equation}
	
	Our correction model has the same structure as the error detection model, but needs a different training for fine-tuning.
	After the error detection training converges, we randomly replace the original AU $X$ with an universal token $ \langle\mathrm{MASK}  \rangle$
	(i.e., the values of $D$ are all $ \langle\mathrm{MASK} \rangle$ in Eq.\ref{eq:mask}) and remove the BCE loss to further train the model to focus on AU correction. 
	We should note that 
	our model can be viewed as a special comparison-based CAPT (similar to Fig.\ref{fig:com_based}). The main differences are: First, we generate AU instead of the spectrum, and AU is more robust to speaker or style variation when used for comparison-based CAPT. Second, the inference of Transformer-based TTS is auto-regressive, as TTS systems only use the text for input. However, for CAPT, both the input speech and its target text are provided. Thus, for inference, we can use the encoded AU for the decoder input. The model performs a partial correction rather than full generation and predicts the end-to-end error.
	\begin{figure}[tb!]
		\centering
		\includegraphics[width=0.75\linewidth]{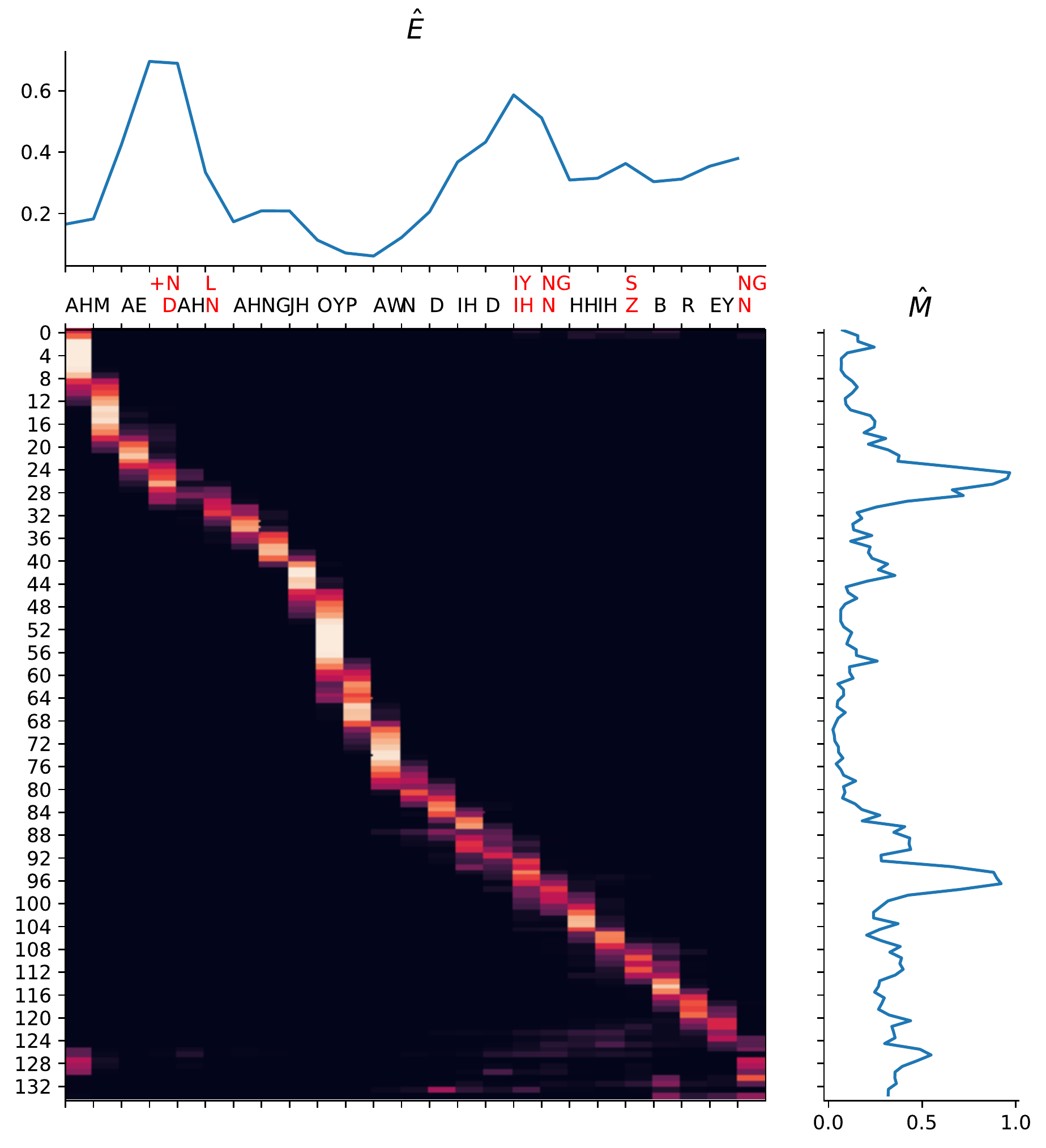}
		\caption{Alignment between the error mask prediction $\hat{M}$ and the phoneme-level error prediction $\hat{E}$. This sample reads ``\textit{A maddening joy pounded in his brain}''. The mispronounced phonemes are marked in red.}
		\label{fig:ali}
	\end{figure}
	
	\subsection{Mispronunciation Correction and Detection}
	Based on the aforementioned training, the detection model decides whether each AU matches the standard L1 distribution conditioned on the input phonemes. Since we use the L1 AU sequence $X$ as the training target in Eq.(\ref{eg:cor}), AU that only appear in L2 utterances or deviated far from the corresponding phoneme will be found out. 

	Note that $\hat{M}$ is the AU-level error prediction. We use the attention map of the last decoder layer to align the AU sequence to the phoneme sequence. Formally, if we define the attention weight between each phoneme $P_i (i\leq L)$ and each predict error mask $\hat{M}_j (j\leq T)$ as $A_{i,j}$, the error prediction for each phoneme $P_i$ is
	\begin{equation}
		\hat{E_i}=\frac{\sum_{j=1}^{T}A_{i,j}\hat{M}_j}{\sum_{j=1}^{T}A_{i,j}}.
	\end{equation}
	A sample result is shown in Fig.\ref{fig:ali}. When $\hat{E_i}$ is larger than the threshold $H$, we mark this phoneme as a mispronunciation. 
	
	For correction, we mask the predicted error area of the original input AU sequence with $ \langle\mathrm{MASK} \rangle$ and forward it using the correction model.
	The corrected AU sequence is further passed to the VQ-VAE decoder to get the spectrum. Finally, the spectrum is converted to the corrected speech waveform by a vocoder.

	\section{EXPERIMENTS}
		\begin{table}[!tb]
		\centering
		\resizebox{0.45\textwidth}{!}{
			\begin{tabular}{ll}
				\textbf{Model}      & \textbf{Description}                                      \\
				\hline
				\textbf{Spectrum}      & $n_{FFT}=2048$, $n_{win}=1200$, $n_{hop}=300$             \\
				\textbf{VQ Encoder }        &                                                           \\
				\hspace{0.1cm} Conformer Enc*3 & $d_{a}=384$, $d_{ff}=1536$, $h=2$, $ks=7$  \\
				\hspace{0.1cm} Conv Layer & $ks=3$, $stride=2$                                        \\
				\hspace{0.1cm} VQ Layer            & $d_q=64$, $V=512$, $\tau=1.0$                                       \\
				\textbf{Code Decoder}          &                                                           \\
				\hspace{0.1cm} TConv Layer  & $ks=3$, $stride=2$                                        \\
				\hspace{0.1cm} Conformer Dec*3 & $d_{a}=384$, $d_{ff}=1536$, $h=2$, $ks=31$ \\
				\textbf{Text Encoder}        &                                                           \\
				\hspace{0.1cm} Conv Layer*2        & $ks=3$, $stride=1$                                        \\
				\hspace{0.1cm} Transformer Enc*6 & $d_{a}=512$, $d_{ff}=1024$,  $h=4$ \\
				\textbf{Code Corrector}      &                                                           \\
				\hspace{0.1cm} Conv Layer*2        & $ks=5$, $stride=1$                                        \\
				\hspace{0.1cm} Transformer Dec*12 & $d_{a}=512$, $d_{ff}=1024$, $h=4$\\
				\hline
			\end{tabular}
		}
			\caption{Model details.}
	\label{tab:model}
	\end{table}
	\label{sec:guidelines}
	\subsection{Implementation Details}
	We use Librispeech\cite{Panayotov41920154242015} as the L1 dataset and L2-Arctic\cite{Zhao2018b} as the L2 dataset. Note that L2-Arctic contains 3599 human-annotated utterances that contain the phoneme-level mispronunciation label, while other 23268 utterances are not annotated. The annotated part is kept as the testset, and the other utterances are combined with Librispeech to train the proposed VQ-VAE. 
	We convert the 24kHz raw waveform into the 80-dim log-Mel spectrum and use the Parallel-WaveGAN vocoder\cite{Yamamoto20191025} for experiments. 
	
	For the proposed model, we list the details in Table \ref{tab:model}, where $d_{a}$ is the attention dim, $d_{ff}$ is the feed-forward dim, $h$ is the number of attention heads, and $ks$ is the kernel size. Note that a  convolutional (Conv) layer is added before the VQ layer for time-domain down-sampling. Correspondingly, a transposed convolutional (TConv) layer is added after the VQ layer for up-sampling. We use Gumbel-Softmax\cite{Jang2016114} as the VQ layer, where $d_q$ is the dim of the VQ codebook embedding, $V$ is the number of codebooks,  $\tau$ is the temperature. We also add two front Conv layers for the text encoder and AU decoder. We set $n=\mathrm{max}(1,\mathrm{int}(\frac{T}{10}))$ for each AU sequence $X_{0:T}$. $k$ and $j$ are uniformly sampled, $k=\mathrm{int}(R_k), R_k\sim\mathcal{U}(0,10)$ and $j=\mathrm{int}(R_j), R_j\sim\mathcal{U}(0,T-k)$. We set $H=0.4$.
	
	We use the Adam optimizer with the warm-up learning scheduler to train both VQ-VAE and the error detection model until the training loss converges. We fine-tune the correction model based on the error detection model with the learning rate of $lr=10^{-4}$.

	\begin{table}[!tb]
		\resizebox{0.43\textwidth}{!}{
		\begin{tabular}{lllll}
			
			\centering
			\textbf{Model}             & \textbf{Dataset}    & \textbf{PRE} & \textbf{REC} & \textbf{F1} \\
			\hline
			\textbf{GOP-Based}         &                     &                    &                 &             \\
			GMM-HMM                    & L1 Only             & 0.290              & 0.290           & 0.290       \\
			\textbf{ASR-Based}         &                     &                    &                 &             \\
			CTC-ATT                    & L1 Only             & 0.305              & 0.525           & 0.386       \\
			Transformer                    & L1 Only             & 0.327              & 0.553           & 0.411       \\
			\textbf{Comparison-Based}  &                     &                    &                 &             \\
			Fastspeech2(w/o Style)     & L1 Only             & 0.190              & 0.527           & 0.279       \\
			Fastspeech2(w/ Style)       & L1+L2(Speech Only) & 0.231              & 0.543           & 0.324       \\
			\textbf{Proposed}          &                     &                    &                 &             \\
			MaskedAU                   & L1+L2(Speech Only) & 0.353              & 0.577           & \textbf{0.438}       \\
			\hline       
			\textbf{ASR-Based Topline} &                     &                    &                 &             \\
			
			AntiPhone          & L1+L2(Supervised)   & 0.499              & 0.613           & 0.550 \\
			\hline      			
		\end{tabular}}
	\caption{Mispronunciation detection results.}
	\label{tab:detection} 
	\end{table}
	
\begin{figure}[t!]
	\centering
	\includegraphics[width=0.75\linewidth]{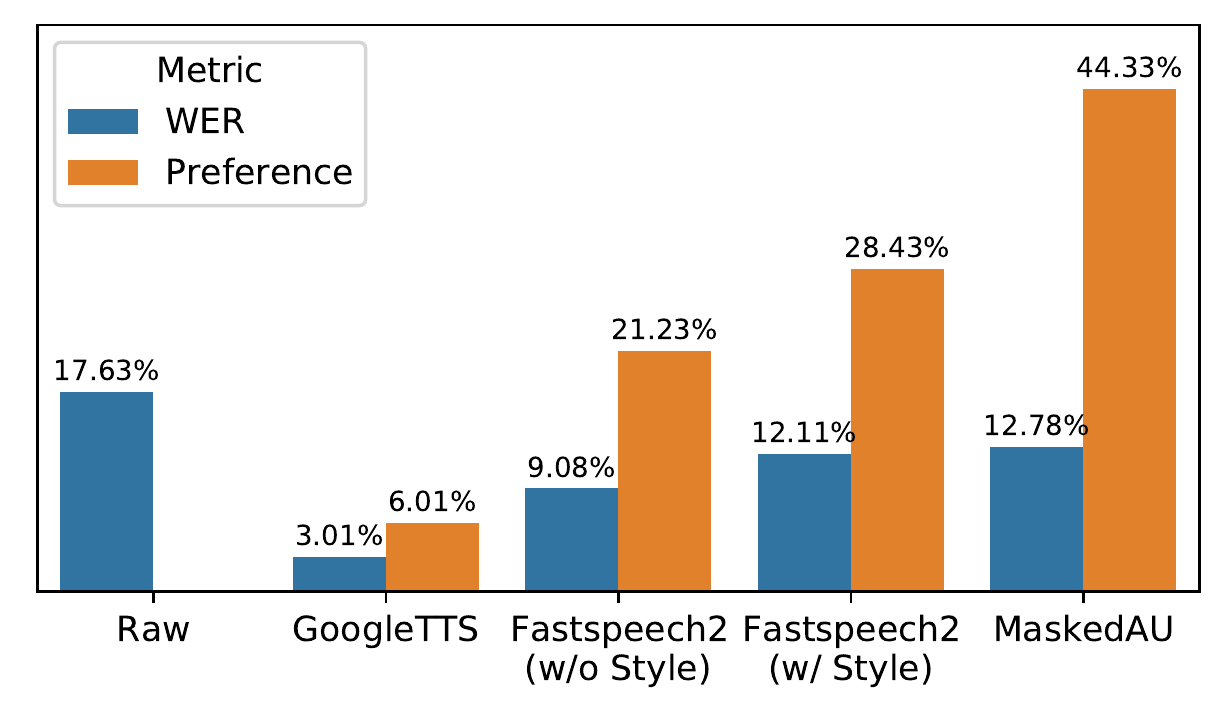}
	\caption{Mispronunciation correction results.}
	\label{fig:correction}
\end{figure}

	\subsection{Mispronunciation Detection}
    For mispronunciation detection, phoneme-level ASR systems trained on Librispeech are set as the baseline using the workflow in \cite{Leung2019,Zhang2020}. Results of the GOP-based GMM-HMM from  \cite{Zhao2018b} are also displayed. For the comparison-based CAPT, we use a TTS model called Fastspeech2\cite{Ren202068}	to generate the standard spectrum. To decrease the error caused by speaker or style variation, we provide the speaker-embedding for generation. Another version also adds the style (energy and pitch from the input L2 utterances) . These two versions are denoted as Fastspeech2 (w/o Style) and Fastspeech2 (w/ Style).	The phoneme duration of the test sample is used as the input condition so that the generated spectrum can be aligned with this sample for error calculation. To get the phoneme-level detection result, we average the frame-level error  between the generated spectrum and the input spectrum for each phoneme. We use the MSE loss or cosine similarity to indicate the error and find MSE is better (the detection threshold is 0.3).
	Finally, we also set a supervised topline called Anti-Phone\cite{Yan2020} that uses the 80\% of the annotated label for training and 20\% for testing.  We use precision (PRE), recall (REC) and F1 score as the metrics.
	
	We show the results in Table \ref{tab:detection}. With the help of deep-learning, the ASR-based models perform better than the GOP-based GMM-HMM model. 
	The comparison-based methods do not perform well for detection as the low-level spectrum error may not be robust enough to overcome the style variation. Adding L2 style alleviates this phenomenon.  Although the topline has the best F1 score, this supervised model must utilize the human-annotated data for training. Thus, it can be expensive and subjective compared to annotation-free models. The proposed model uses the robust AU sequence and the conditioned phonemes for judgment and achieves the best performance among the annotation-free models.

	\subsection{Mispronunciation Correction}
	To test whether the generated speech is corrected to the standard one objectively, we use a word-level ASR system trained on Librispeech\footnote{\url{https://huggingface.co/speechbrain/asr-crdnn-transformerlm-librispeech}.} to test the word error rate (WER). A higher WER suggests that the generated speech contains more mispronunciations that cannot be recognized by this L1 ASR system. We use the former two comparison-based models as the baseline. WER of the raw speech and the speech generated by GoogleTTS are also displayed. 
	As WER is also limited by the used ASR system, we provide the input and the prediction and ask 20 volunteers to vote their preference of using each model for CAPT. 
	
	As we can see from Fig.\ref{fig:correction}, although the standard speech generated by GoogleTTS has the lowest WER, this model can only offer a fixed speech for teaching. Thus, its preference is relatively low. The other methods can partially clone the voice of the speaker, making  a self-intimating feedback. They are more interesting and powerful for education. Adding styles from the L2 input leads to an increased WER but the preference also increases. For our method, as we only modify the wrong AU and keep the correct AU, it achieves the best style preservation for self-intimating. The results show that the proposed method achieves a comparable WER performance and the best preference. 
	
	\section{Conclusion}
	In this paper, we propose a novel method to utilize AU for both mispronunciation detection and correction. Compared with ASR-based CAPT, our method can achieve a good detection performance without using the expensive annotation. Compared with comparison-based CAPT, our method is more robust in detection and also performs well in correction.
	Experiments show that the proposed method is a promising approach for CAPT. 

	\newpage
	\bibliographystyle{IEEEbib}
	\bibliography{refs}

\begin{thebibliography}{10}

\bibitem{Witt2000}
S.~M. Witt and S.~J. Young,
\newblock ``Phone-level pronunciation scoring and assessment for interactive
  language learning,''
\newblock {\em Speech Communication}, vol. 30, no. 2, pp. 95--108, 2000.

\bibitem{Zhang2020}
Long Zhang, Ziping Zhao, Chunmei Ma, Linlin Shan, Huazhi Sun, Lifen Jiang,
  Shiwen Deng, and Chang Gao,
\newblock ``End-to-end automatic pronunciation error detection based on
  improved hybrid ctc/attention architecture,''
\newblock {\em Sensors (Switzerland)}, vol. 20, no. 7, pp. 1--24, 2020.

\bibitem{Yan2020}
Bi~Cheng Yan, Meng~Che Wu, Hsiao~Tsung Hung, and Berlin Chen,
\newblock ``An end-to-end mispronunciation detection system for l2 english
  speech leveraging novel anti-phone modeling,''
\newblock in {\em Proceedings of the Annual Conference of the International
  Speech Communication Association, INTERSPEECH}, 2020, pp. 3032--3036.

\bibitem{Zhang2021}
Zhan Zhang, Yuehai Wang, and Jianyi Yang,
\newblock ``Text-conditioned transformer for automatic pronunciation error
  detection,''
\newblock {\em Speech Communication}, vol. 130, pp. 55--63, 2021.

\bibitem{Leung2019}
Wai-Kim Leung, Xunying Liu, and Helen Meng,
\newblock ``Cnn-rnn-ctc based end-to-end mispronunciation detection and
  diagnosis,''
\newblock in {\em IEEE International Conference on Acoustics, Speech and Signal
  Processing, ICASSP}, 2019.

\bibitem{Chang2010}
Charles~Bond Chang,
\newblock ``First language phonetic drift during second language acquisition,''
\newblock {\em ProQuest LLC}, 2010.

\bibitem{Tu2018}
Ming Tu, Anna Grabek, Julie Liss, and Visar Berisha,
\newblock ``Investigating the role of l1 in automatic pronunciation evaluation
  of l2 speech,''
\newblock in {\em Proceedings of the Annual Conference of the International
  Speech Communication Association, INTERSPEECH}, 2018, pp. 1636--1640.

\bibitem{Baevski2020620}
Alexei Baevski, Yuhao Zhou, Abdelrahman Mohamed, and Michael Auli,
\newblock ``wav2vec 2.0: A framework for self-supervised learning of speech
  representations,''
\newblock {\em Advances in Neural Information Processing Systems}, vol. 33,
  2020.

\bibitem{Hsu9414460}
Wei-Ning Hsu, Yao-Hung~Hubert Tsai, Benjamin Bolte, Ruslan Salakhutdinov, and
  Abdelrahman Mohamed,
\newblock ``Hubert: How much can a bad teacher benefit asr pre-training?,''
\newblock in {\em IEEE International Conference on Acoustics, Speech and Signal
  Processing, ICASSP}, 2021, pp. 6533--6537.

\bibitem{vanNiekerk2020}
Benjamin {van Niekerk}, Leanne Nortje, and Herman Kamper,
\newblock ``Vector-quantized neural networks for acoustic unit discovery in the
  zerospeech 2020 challenge,''
\newblock in {\em Proceedings of the Annual Conference of the International
  Speech Communication Association, INTERSPEECH}, 2020, pp. 4836--4840.

\bibitem{Lee12220121252012}
Ann Lee and James Glass,
\newblock ``A comparison-based approach to mispronunciation detection,''
\newblock in {\em IEEE Spoken Language Technology Workshop, SLT}, 2012, pp.
  382--387.

\bibitem{Yang1519September2019}
Seung~Hee Yang and Minhwa Chung,
\newblock ``Self-imitating feedback generation using gan for computer-assisted
  pronunciation training,''
\newblock in {\em Proceedings of the Annual Conference of the International
  Speech Communication Association, INTERSPEECH}, 2019, pp. 1881--1885.

\bibitem{Gulati2020}
Anmol Gulati, James Qin, Chung-Cheng Chiu, Niki Parmar, Yu~Zhang, Jiahui Yu,
  Wei Han, Shibo Wang, Zhengdong Zhang, Yonghui Wu, and Ruoming Pang,
\newblock ``Conformer: Convolution-augmented transformer for speech
  recognition,''
\newblock in {\em Proceedings of the Annual Conference of the International
  Speech Communication Association, INTERSPEECH}, 2020.

\bibitem{Vaswani2017}
Ashish Vaswani, Noam Shazeer, Niki Parmar, Jakob Uszkoreit, Llion Jones,
  Aidan~N. Gomez, {\L}ukasz Kaiser, and Illia Polosukhin,
\newblock ``Attention is all you need,''
\newblock {\em Advances in Neural Information Processing Systems}, vol.
  2017-Decem, pp. 5999--6009, 2017.

\bibitem{Panayotov41920154242015}
Vassil Panayotov, Guoguo Chen, Daniel Povey, and Sanjeev Khudanpur,
\newblock ``Librispeech: An asr corpus based on public domain audio books,''
\newblock in {\em IEEE International Conference on Acoustics, Speech and Signal
  Processing, ICASSP}, 2015, pp. 5206--5210.

\bibitem{Zhao2018b}
Guanlong Zhao, Sinem Sonsaat, Alif Silpachai, Ivana Lucic, Evgeny
  Chukharev-Hudilainen, John Levis, and Ricardo Gutierrez-Osuna,
\newblock ``L2-arctic: A non-native english speech corpus,''
\newblock in {\em Proceedings of the Annual Conference of the International
  Speech Communication Association, INTERSPEECH}, 2018.

\bibitem{Jang2016114}
Eric Jang, Shixiang Gu, and Ben Poole,
\newblock ``Categorical reparameterization with gumbel-softmax,''
\newblock {\em arXiv preprint arXiv:1611.01144}, 2016.

\bibitem{Ren202068}
Yi~Ren, Chenxu Hu, Xu~Tan, Tao Qin, Sheng Zhao, Zhou Zhao, and Tie-Yan Liu,
\newblock ``Fastspeech 2: Fast and high-quality end-to-end text to speech,''
\newblock {\em arXiv preprint arXiv:2006.04558}, 2020.

\bibitem{Yamamoto20191025}
Ryuichi Yamamoto, Eunwoo Song, and Jae-Min Kim,
\newblock ``Parallel wavegan: A fast waveform generation model based on
  generative adversarial networks with multi-resolution spectrogram,''
\newblock {\em arXiv preprint arXiv:1910.11480}, 2019.

\end{thebibliography}
\end{document}